\documentclass[aps,prl,twocolumn,amsmath,amssymb,nofootinbib,superscriptaddress]{revtex4-1}

\usepackage{graphicx}
\usepackage{dcolumn}
\usepackage{bm}
\usepackage{amsmath}
\usepackage{indentfirst}
\usepackage{float}
\usepackage[colorlinks]{hyperref}
\usepackage[dvipsnames]{xcolor}

\begin{document}

\title{Experimental Blind Quantum Computing for a Classical Client}

\author{He-Liang Huang}
\affiliation{Hefei National Laboratory for Physical Sciences at Microscale and Department of Modern Physics,\\
University of Science and Technology of China, Hefei, Anhui 230026, China}
\affiliation{CAS-Alibaba Quantum Computing Laboratory, CAS Centre for Excellence in Quantum Information and Quantum Physics, University of Science and Technology of China, Shanghai 201315, China}
\affiliation{Henan Key Laboratory of Quantum Information and Cryptography, Zhengzhou Information Science and Technology Institute, Henan, Zhengzhou 450000, China}
\author{Qi Zhao}
\affiliation{Center for Quantum Information, Institute for Interdisciplinary Information Sciences, Tsinghua University, Beijing
100084, China}
\author{Xiongfeng Ma}
\affiliation{Center for Quantum Information, Institute for Interdisciplinary Information Sciences, Tsinghua University, Beijing
100084, China}
\author{Chang Liu}
\author{Zu-En Su}
\affiliation{Hefei National Laboratory for Physical Sciences at Microscale and Department of Modern Physics,\\
University of Science and Technology of China, Hefei, Anhui 230026, China}
\affiliation{CAS-Alibaba Quantum Computing Laboratory, CAS Centre for Excellence in Quantum Information and Quantum Physics, University of Science and Technology of China, Shanghai 201315, China}
\author{Xi-Lin Wang}
\author{Li Li}
\author{Nai-Le Liu}
\affiliation{Hefei National Laboratory for Physical Sciences at Microscale and Department of Modern Physics,\\
University of Science and Technology of China, Hefei, Anhui 230026, China}
\affiliation{CAS-Alibaba Quantum Computing Laboratory, CAS Centre for Excellence in Quantum Information and Quantum Physics, University of Science and Technology of China, Shanghai 201315, China}
\author{Barry C. Sanders}
\affiliation{Hefei National Laboratory for Physical Sciences at Microscale and Department of Modern Physics,\\
University of Science and Technology of China, Hefei, Anhui 230026, China}
\affiliation{CAS-Alibaba Quantum Computing Laboratory, CAS Centre for Excellence in Quantum Information and Quantum Physics, University of Science and Technology of China, Shanghai 201315, China}
\affiliation{Institute for Quantum Science and Technology, University of Calgary, Alberta T2N 1N4, Canada}
\affiliation{Program in Quantum Information Science, Canadian Institute for Advanced Research, Toronto, Ontario M5G 1Z8, Canada}
\author{Chao-Yang Lu}
\author{Jian-Wei Pan}
\affiliation{Hefei National Laboratory for Physical Sciences at Microscale and Department of Modern Physics,\\
University of Science and Technology of China, Hefei, Anhui 230026, China}
\affiliation{CAS-Alibaba Quantum Computing Laboratory, CAS Centre for Excellence in Quantum Information and Quantum Physics, University of Science and Technology of China, Shanghai 201315, China}

\date{\today}


\begin{abstract}
To date, blind quantum computing demonstrations require clients to have weak quantum devices. Here we implement a proof-of-principle experiment for completely classical clients. Via classically interacting with two quantum servers that share entanglement, the client accomplishes the task of having the number 15 factorized by servers who are denied information about the computation itself. This concealment is accompanied by a verification protocol that tests servers' honesty and correctness. Our demonstration shows the feasibility of completely classical clients and thus is a key milestone towards secure cloud quantum computing.
\end{abstract}

\maketitle
Whereas quantum computers could exponentially outperform classical computers for certain computational tasks, inaccessibility due to implementation complexity would hinder widespread adoption of quantum computing. Thus, quantum computation is increasingly being performed `in the cloud', such as IBM's 5-qubit quantum cloud service \cite{IBMQE}. In this approach, quantum computing is outsourced from a client with classical hardware to a server who possesses expensive quantum hardware. Considering the types of applications to which quantum computing is likely to be applied, imformation security is important as clients may wish to keep the computation perfectly secret from  untrusted servers implementing the quantum computation.

A solution to this issue is offered by blind quantum computing (BQC) \cite{broadbent2009universal}, which is a quantum cryptographic protocol that enables a classical client with limited quantum technology to delegate a computation to the quantum server(s) without leaking any information about her computation to the server(s). Thus far various BQC protocols have been proposed \cite{broadbent2009universal,fitzsimons2012unconditionally,aharonov2008interactive,morimae2012blind,morimae2013blind,morimae2014verification,hayashi2015verifiable,dunjko2012blind,hajduvsek2015device,gheorghiu2015robustness,mantri2013optimal,perez2015iterated,
reichardt2013classical,gheorghiu2015rigidity} , and some proof-of-principle experiments have been performed with photonic qubits \cite{barz2012demonstration,barz2013experimental,fisher2014quantum,greganti2016demonstration,marshall2016continuous}. However, all these experimental demonstrations only support quasi-classical clients. That is, the clients require the ability to prepare or measure single-qubit states, but wide use of quantum computing on the cloud would be much more attractive if clients did not require the ability to perform quantum tasks. Although using only classical communication between a classical client and a \emph{single} quantum server may be infeasible for secure BQC~\cite{aaronson2017implausibility}, classical communication between a classical client and multi-quantum servers can work \cite{reichardt2013classical}.

Besides security, verifiability is another important concern for BQC, i.e.\ the ability of a client to test whether or not the servers perform the task correctly and honestly. As the complexity of quantum many-body systems scales up, verifiability becomes a major experimental challenge, not only in BQC, but also in quantum chemistry \cite{gan2005calibrating}, quantum simulation \cite{senko2014coherent}, BosonSampling \cite{aaronson2014bosonsampling}, and other quantum algorithms. Thus, a verification protocol for BQC is significant not only as a cryptographic protocol but also for exploring the relation between quantum physics and computer science.

Here we demonstrate a proof-of-principle implementation of BQC for completely classical clients. In our experiment, we realize Shor's algorithm \cite{shor1999polynomial} for factorizing ${N = 15}$ via the framework of verifiable BQC based on the Reichardt, Unger and Vazirani (RUV) protocol \cite{reichardt2013classical}. The scheme employs quantum gate teleportation for computation and combines the rigidity of Clauser-Horne-Shimony-Holt (CHSH) tests \cite{reichardt2013classical} and stabilizer tests for verification, thereby providing a method for a client to control quantum servers classically.

Suppose we are given two quantum servers, Alice and Bob, that share Einstein-Podolsky-Rosen (EPR) states but cannot communicate with each other (enforced e.g.\ through space-like separation of the devices). The client Charlie, holding a completely classical device, wants to delegate quantum computing to the remote servers without leaking any imformation about the computation to servers. He can decompose the circuit into two parts, Computation-A and Computation-B, and send these two tasks to Alice and Bob, respectively. Alice and Bob operate on their respective halves of the shared EPR states according to Charlie's commands and return to Charlie the measurement results. As Alice and Bob cannot communicate with each other, they cannot learn the results from each other, so this delegated computation is `blind', meaning that each server learns nothing more about the computation than its length \cite{reichardt2013classical}.

\begin{figure}
\includegraphics[width=\columnwidth]{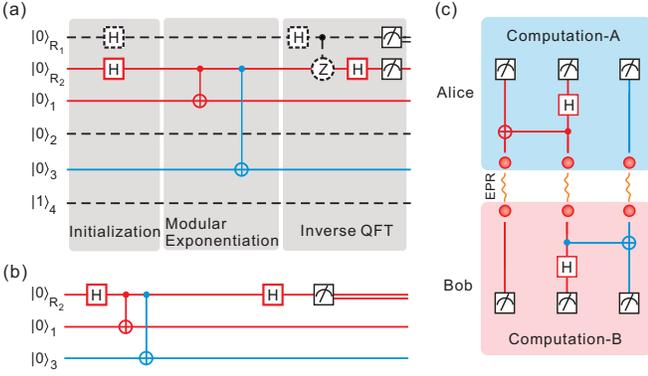}
\caption{(color online). Quantum circuit for factorizing ${N = 15}$ using Shor's algorithm. (a) Quantum circuit for ${N = 15}$ and ${a = 11}$ \cite{lu2007demonstration}. The modular exponential function is implemented by two CNOT gates, and the QFT is implemented by Hadamard rotations and two-qubit conditional-phase gates. (b) The simplified version of the circuit in (a), omitting the qubits and operations marked by dotted lines in (a). (c) The scheme of cloud quantum computing for factorizing ${N = 15}$. Each measurement is in the $Z$ basis.} \label{fig:1}
\end{figure}

For the task of factorizing ${N}$ using Shor's algorithm, if we pick a random number ${a}$ that is co-prime to $N$, Shor's algorithm can yield the minimum integer $r$ that satisfies ${{a^r}{\kern 1pt} \bmod {\kern 1pt} N = 1}$.  From this period $r$, the prime factors of $N$ are given by the greatest common divisor (GCD) of ${{a^{r/2}}{\kern 1pt}  \pm 1}$ and ${N}$, which is solved classically. The quantum circuit for $N=15$ and ${a = 11}$ is shown in Fig.\ \ref{fig:1}(a) \cite{lu2007demonstration}. In fact, The inverse quantum Fourier transformation (QFT) is unnecessary for any order-${{2^l}}$ circuit \cite{lanyon2007experimental}. Moreover, two qubits ${|0{\rangle _2}}$ and ${|1{\rangle _4}}$ evolve trivially during the computation and thus can be omitted. This fact allows us to simplify the circuit to Fig.\ \ref{fig:1}(b) by omitting obsolete qubits and operations marked by dotted lines in the circuit in Fig.\ \ref{fig:1}(a).

To delegate the circuit in Fig.\ \ref{fig:1}(b) to two remote quantum servers, Charlie decomposes it into two parts (see Fig.\ \ref{fig:1}(c)) and sends the tasks to Alice and Bob, respectively.  Each observable of Alice (Bob) has eigenvalues ${ \pm 1}$ such that each outcome ${{a_i}{\rm{ }}({b_i})}$ reported to Charlie takes values of ${ \pm 1}$, where ${i}$ denotes the $i$th qubit of Alice (Bob). By design, Computation-A performs the first controlled-NOT (CNOT) gate of the circuit and prepares the third input state ${|0\rangle }$ for Bob. If Alice implements Computation-A honestly, Bob's share of EPR states collapses into ${|\varPsi \rangle |\beta \rangle }$, where ${|\varPsi \rangle }$ is one of the four Bell states, and ${|\beta \rangle  \in \{ |0\rangle ,|1\rangle \} }$, according to Alice's results. In particular, when Alice reports ${{a_1} = {a_2} = {a_3} = 1}$, Bob's state collapses into the desired resource state ${|\phi \rangle  = |\Phi ^ +\rangle |0\rangle }$, where ${|\Phi ^ \pm\rangle = \frac{1}{{\sqrt 2 }}(|00\rangle  \pm |11\rangle)}$, which is equivalent to the state after the first CNOT gate in the circuit in Fig.\ \ref{fig:1}(b). Then Bob implements Computation-B to achieve the second CNOT gate and measures his 2nd qubit in the Pauli $X$ basis to output the result of Shor's algorithm. Bob's remaining two qubits contribute nothing to the outcome and are both measured in the Pauli $Z$ basis as they can be employed in the validation procedure described below.

When performing such a computation on untrusted quantum servers, clients also wish to test the honesty of servers: did they implement the computation as promised? To realize this test, Charlie randomly switch tasks being implemented by Alice and Bob between the desired computation and `dummy' protocols. The dummy protocols are constructed such that Alice and Bob are unable to distinguish whether they are implementing the proper computation or the dummy, but such that Charlie is able to detect if the dummy tasks are being implemented dishonestly. Via repeated application of this randomized procedure, Charlie then determines whether Alice and Bob are being honest. Specifically, Charlie can randomly command the servers (see Fig.\ \ref{fig:2}(a)) to implement the four sub-protocols below:

\begin{figure*}
\includegraphics[width=1.4\columnwidth]{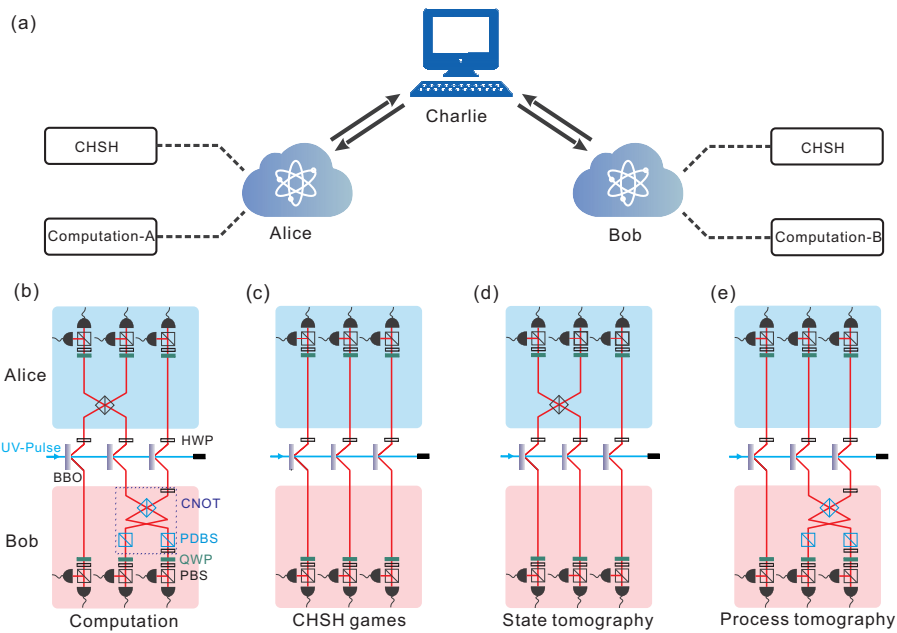}
\caption{(color online). Experimental setup. (a) Outline of the scheme. Charlie classically interacts with quantum servers Alice and Bob who share entanglement. Each of the quantum servers is randomly commanded to implement one of the two types of operations, CHSH and Computation-A(B). (b) Computation setup. Ultraviolet laser pulses with a central wavelength of 394 nm, pulse duration of 150fs, and repetition rate of 80MHz pass through three BBO crystals to produce three polarization-entangled pairs ${\frac{1}{{\sqrt 2 }}\left( {|H\rangle |V\rangle  + |V\rangle |H\rangle } \right)}$. A HWP is placed at an arm of the entangled pairs to produce EPR states ${\frac{1}{{\sqrt 2 }}\left( {|H\rangle |H\rangle  + |V\rangle |V\rangle } \right)}$. To achieve good spatial and temporal overlap, all photons are spectrally filtered with 3-nm bandwidth filters. The final measurement results are then read out by single-photon detectors with dual-channel structure, which partially eliminates higher-order events. (c) CHSH test setup. (d) State tomography setup. (e) Process tomography setup.} \label{fig:2}
\end{figure*}

1. Computation. As shown in Fig.\ \ref{fig:1}(c), the computation is realized as the joint evolution of two isolated quantum servers. In our experiment, Computation-A and Computation-B can be compiled into the setup in Fig.\ \ref{fig:2}(b), where the logical qubits ${|0\rangle }$ and ${|1\rangle }$ are encoded by horizontal ($H$) and vertical ($V$) polarizations of single photons, respectively. Instead of implementing the standard CNOT gate between the first and second qubits in Computation-A, Charlie can ask Alice to use a polarizing beam splitter (PBS) to postselect events where there is exactly one photon exiting each output (the first two EPR states are transformed into ${\rm{1/}}\sqrt 2 {\rm{(|}}0\rangle _{\rm{1}}^{\rm{A}}|0\rangle _{\rm{2}}^{\rm{A}}{\rm{|}}0\rangle _{\rm{1}}^{\rm{B}}|0\rangle _{\rm{2}}^{\rm{B}} + {\rm{|}}1\rangle _{\rm{1}}^{\rm{A}}|1\rangle _{\rm{2}}^{\rm{A}}{\rm{|}}1\rangle _{\rm{1}}^{\rm{B}}|1\rangle _{\rm{2}}^{\rm{B}})$ after postselection, where A (B) represents Alice (Bob).) and measure these two photons in the Pauli ${X}$ basis.

If Alice's reported results yield ${{a_1} \cdot {a_2} = {a_3} = 1}$, then Bob's share of the EPR states collapse onto the desired state ${|\phi \rangle }$. The CNOT gate in Computation-B can be realized by combining three polarization-dependent beamsplitters (PDBS) -- an overlapping PDBS (${{T_H} = 1}$ and ${{T_V} = 1/3}$), and two supplementary PDBSs (${{T_V} = 1}$ and ${{T_H} = 1/3}$) at each exit port of the overlapping PDBS, along with two Hadamard gates (half-wave plate (HWP)) on the target photon before and after the PDBS \cite{kiesel2005linear}. The different treatment of the CNOT gates arises because Bob is required to complete the computation and convey the final outcomes, so he is instructed to implement the complete Bell measurement. However, Alice only needs to prepare resource states for Bob. As long as Alice can prepare the desired states, we deem her to be honest.

2. CHSH test. Charlie sends random bits ${A \in \{ 0,1\} }$ and ${B \in \{ 0,1\} }$ to Alice and Bob, respectively, which determines their measurement bases, and they respond with bits ${M \in \{ 0,1\} }$ and ${N \in \{ 0,1\} }$ corresponding to their binary measurement outcomes (see Fig.\ \ref{fig:2}(c)). In this test, Alice and Bob `win' if ${AB = M \oplus N}$, and they can win with probability ${{\omega ^ * } = {\cos ^2}(\pi /8) \approx 0.854}$ if Bob measures in the Pauli ${Z}$ basis for $B=0$ or Pauli ${X}$ basis for $B=1$, and if Alice measures ${(Z + {( - 1)^A}X)/\sqrt 2 }$. According to Alice's and Bob's measurement outcomes $a$ and $b$, i.e., $\pm$ 1, Charlie sets $M$ and $N$ to 0 or 1. In contrast, classical servers can win with probability at most ${3/4}$. In our protocol, Charlie can also change the strategy to simultaneously swap the measurement bases of Alice and Bob; that is, Alice measures ${Z}$ or ${X}$, and Bob measures ${(Z + {( - 1)^B}X)/\sqrt 2 }$. According to the rigidity of CHSH test \cite{reichardt2013classical}, if the servers win with probability close to ${\omega ^ * }$, the implement strategy is close to the ideal strategy. To ensure servers' honesty, Charlie runs $n$ rounds of CHSH tests with both servers, and rejects if the servers fail to win at least ${(\omega ^ * - \varepsilon )n}$ rounds, where ${\varepsilon=\frac{1}{2\sqrt{2}}\sqrt{\log{n}/n}}$ is the error threshold \cite{reichardt2012classical,Supplemental}.

3. State tomography. Charlie asks Alice to implement Computation-A while running CHSH test with Bob (see Fig.\ \ref{fig:2}(d)).  If Alice honestly implements the command, Bob's state collapses to ${|\Phi ^ \pm\rangle \otimes |\beta \rangle }$. Bob is required to measure in the bases ${{X_1}{X_2}{Z_3}}$ or ${{Z_1}{Z_2}{Z_3}}$, where the first two bases ${{X_1}{X_2}}$ and ${{Z_1}{Z_2}}$ are the stabilizers for the Bell states, and ${{Z_3}}$ is the stabilizer of ${|\beta \rangle }$. In these cases, Bob's measurement outcomes are deterministic, depending on Alice's results. Thus, Charlie can test whether Alice is honest according Bob's measurement outcomes. If Bob reports the wrong stabilizer syndrome in even a single round, Charlie can reject. If Alice plays honestly, Charlie accepts with high probability.

\begin{figure*}
\includegraphics[width=1.42\columnwidth]{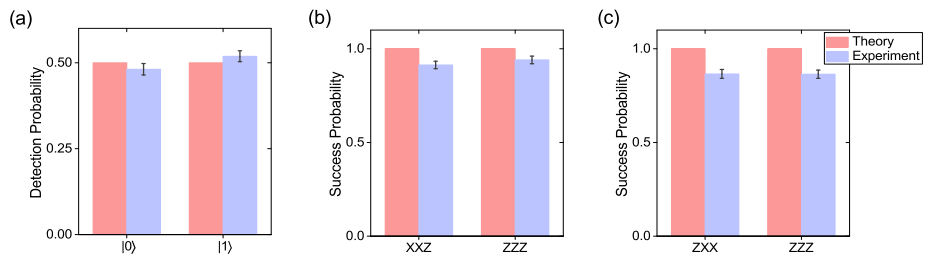}
\caption{(color online). Experimental results for honest Alice and Bob. (a) Output of quantum computing for factorizing ${N = 15}$, which is determined by the results of the second photon Bob observed in the subprotocol Computation. Theoretical predictions and measured expectation values are shown as red and blue bars, respectively. (b) The probability that Alice passes the tests of state tomography when Bob measures in the ${{X_1}{X_2}{Z_3}}$ and ${{Z_1}{Z_2}{Z_3}}$ bases. (c) The probability that Bob passes the tests of process tomography when Alice measures in the ${{Z_1}{X_2}{X_3}}$ and ${{Z_1}{Z_2}{Z_3}}$ bases.} \label{fig:3}
\end{figure*}

4. Process tomography. Charlie asks Bob to implement Computation-B while running CHSH test with Alice (see Fig.\ \ref{fig:2}(e)). If Bob honestly implements the command, Alice's state collapses to ${|\beta \rangle |\varPsi \rangle }$. Alice is required to measure in the bases ${{Z_1}{X_2}{X_3}}$ or ${{Z_1}{Z_2}{Z_3}}$, where the last two bases, ${{X_2}{X_3}}$ and ${{Z_2}{Z_3}}$, are the Bell-state stabilizers, and the first basis ${{Z_1}}$ is the stabilizer of ${|\beta \rangle }$. Therefore, if Alice reports the wrong stabilizer syndrome in even a single round, Charlie can reject. If Bob plays honestly, Charlie accepts with high probability.

Charlie runs Protocol 1 with a small probability $\eta$ , and other three alternative protocols with probability $\frac{1-\eta}{3}$ so that servers are not aware of which protocol their measurements belong to. For instance, from Alice's perspective, she is entirely unaware whether Bob is implementing CHSH test or Computation-B. From the CHSH test and stabilizer test, Charlie can determine whether the servers are being honest or not. The relationship among $\eta$, computational efficiency and security parameters are analyzed in Supplemental Material \cite{Supplemental}.

To demonstrate the scheme, we employ polarization-entangled photons ${|\Phi ^ + \rangle}$ generated by spontaneous parametric down-conversion using a HWP-sandwiched $\beta$-barium borate (BBO) crystal \cite{PhysRevLett.117.210502}. For Protocol 1, experimental results are shown in Fig.\ \ref{fig:3}(a). If Alice and Bob play honestly, then, with probability  ${ \sim {\rm{51}}{\rm{.9\% }}}$, the output is ${|0\rangle }$, corresponding to a failure. The remaining ${ \sim {\rm{48}}{\rm{.1\% }}}$ probability yields ${|1\rangle }$. Combining these with the known state of the redundant qubit ${|0{\rangle _{{R_1}}}}$ using classical processing yields the period ${r = 2}$. Thus, ${{\rm{GCD}}({11^{2/2}} \pm 1,15) = 3,5}$, yielding a successful factorization. To quantify the performance of the CNOT operations realized by the PDBS, we measure process fidelity \cite{hofmann2005complementary} for the CNOT gate as ${0.87(2) \le {F_{{\rm{process}}}} \le 0.93(2)}$ (see Supplemental Material \cite{Supplemental} for details).

In our experiment, we run $n=6000$ rounds of CHSH tests, then the error threshold is calculated as ${\varepsilon  = \frac{1}{2\sqrt{2}}\sqrt{\log n/n}=0.014}$. Two honest quantum servers win with the probability $\sim$0.846(6), from which ${\varepsilon }$ is calculated as ${\varepsilon  =}$0.007(6) -- below the error threshold. Thus, Charlie accepts the protocol (see Supplemental Material \cite{Supplemental} for more detailed security analysis). On the other hand, if the quantum servers play dishonestly, for example, making the angle of the HWP in Bob's measurement set-up always ${{5^ \circ }}$ higher than the target angle, they win with probability $\sim$0.814(5) and thus ${\varepsilon  =}$0.047(5), which is above the threshold, and Charlie rejects.

\begin{figure}
\includegraphics[width=0.92\columnwidth]{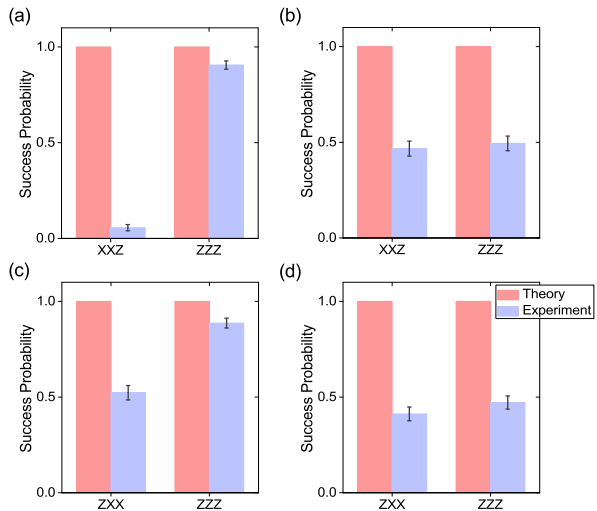}
\caption{(color online). Experimental result for dishonest Alice and Bob. (a) Probability that Alice passes the tests of state tomography when Bob measures in ${{X_1}{X_2}{Z_3}}$ and ${{Z_1}{Z_2}{Z_3}}$ bases, if Alice deliberately reports the opposite results for the first qubit. (b) Probability that Alice passes the tests of state tomography when Bob measures in ${{X_1}{X_2}{Z_3}}$ and ${{Z_1}{Z_2}{Z_3}}$ bases, if Alice's third measurement basis is ${X_3}$ instead of ${Z_3}$. (c) Probability that Bob passes the tests of process tomography when Alice measures in ${{Z_1}{X_2}{X_3}}$ and ${{Z_1}{Z_2}{Z_3}}$ bases, if Bob measures the last two qubits in ${{Z_2}{Z_3}}$ instead of the ${{X_2}{Z_3}}$ basis. (d) Probability that Bob passes the tests of process tomography when Alice measures in ${{Z_1}{X_2}{X_3}}$ and ${{Z_1}{Z_2}{Z_3}}$ bases, if Bob's first measurement basis is ${Z_1}$ instead of ${X_1}$.} \label{fig:4}
\end{figure}

Protocol 3 is designed to monitor whether Alice honestly executes Computation-A. If Alice plays honestly, Bob's measurement outcomes are deterministic, depending on Alice's results. Figure~\ref{fig:3}(b) shows the theoretical and experimental results. The probability that Alice passes the tests are 0.92(2) and 0.94(2) when Bob measures in the ${{X_1}{X_2}{Z_3}}$ and ${{Z_1}{Z_2}{Z_3}}$ bases (see Supplemental Material \cite{Supplemental} for details), respectively. To illustrate that the method can detect whether Alice is cheating, we consider two typical potential means of cheating: (1) If Alice deliberately reports the opposite outcomes of the first qubit, and the reported results yield ${{a_1} \cdot {a_2} = 1( - 1)}$, Bob's share of the EPR state collapses into ${|\Phi ^ - \rangle|0\rangle}$ (${|\Phi ^ + \rangle|0\rangle}$) instead of into ${|\Phi ^ + \rangle|0\rangle}$ (${|\Phi ^ - \rangle|0\rangle}$) so the probability of passing the tests (see Fig.\ \ref{fig:4}(a)) drops to 0.06(2) when Bob measures in the ${{X_1}{X_2}{Z_3}}$ basis and remains at 0.91(2) in the ${{Z_1}{Z_2}{Z_3}}$ basis; (2) If Alice's third measurement basis is ${{X_3}}$ instead of ${{Z_3}}$, the probability of passing the tests for ${{X_1}{X_2}{Z_3}}$ and ${{Z_1}{Z_2}{Z_3}}$ measurements are 0.47(4) and 0.49(4) (Fig.\ \ref{fig:4}(b)), respectively. Obviously, Charlie can easily identify that Alice is dishonest based on Bob's reported results.

Protocol 4 monitors whether Bob honestly executes Computation-B. If Bob plays honestly, Alice's measurement outcomes are deterministic, depending on Bob's results. Figure~\ref{fig:3}(c) shows the theoretical and experimental results. The probability that Bob passes the tests are 0.87(2) and 0.86(2) when Alice measures in the ${{Z_1}{X_2}{X_3}}$ and ${{Z_1}{Z_2}{Z_3}}$ bases (see Supplemental Material \cite{Supplemental} for details), respectively. To demonstrate that the method detects whether Bob is cheating, we consider two possible circumstances: (1) If Bob measures the last two qubits in the ${{Z_2}{Z_3}}$ basis instead of the ${{X_2}{Z_3}}$ basis, the probability of passing the tests (Fig.\ \ref{fig:4}(c)) drops to 0.52(4) when Alice measures in the ${{Z_1}{X_2}{X_3}}$ basis and remains at 0.89(3) in the ${{Z_1}{Z_2}{Z_3}}$ basis. (2). If Bob's first measurement basis is ${{Z_1}}$ instead of ${{X_1}}$, the probability of passing ${{Z_1}{X_2}{X_3}}$ and ${{Z_1}{Z_2}{Z_3}}$ tests are 0.41(4) and 0.47(3) (Fig.\ \ref{fig:4}(d)), respectively. Thus, Bob's cheating can easily be caught.

This scheme is device-independent, in that it mitigates the need for clients to place trust in any pre-existing device. The scheme is theoretically efficient, in the sense that its number of rounds scales with  circuit size $n$, $O(n^c)$, where $c$ is a constant \cite{reichardt2013classical, gheorghiu2015robustness}. Subsequent results indicate the number of rounds  can be reduced if we require only one-sided device-independence \cite{gheorghiu2015rigidity}.

In summary, we experimentally demonstrate secure computation on quantum cloud servers using a photonic setup where 3 EPR states are shared between two quantum servers. In our implementation, the correctness of results can be tested through verification protocols, based on the rigidity of CHSH test and stabilizer tests. Our experiment introduces the features of multiple servers, device independence, and, especially, a completely classical client, leading to a heuristic exploration for future secure distributed quantum networks `in the cloud'. This type of encryption will be crucial to enable scalable models for secure, outsourced quantum computation to emerge, paving the way for the commercialization and widespread adoption of quantum computer technology.

We thank Peter P. Rohde, Matthew McKague, Xingyao Wu, Yi-Han Luo, Ching-Yi Lai, Kai-Min Chung, and Xiao Yuan for helpful discussions. This work was supported by the National Natural Science Foundation of China, the Chinese Academy of Sciences, and the National Fundamental Research Program. B.C.S. acknowledges financial support from AITF, NSERC and China's 1000 Talent Plan.
\bibliography{CloudComputation}

\begin{thebibliography}{32}%
\makeatletter
\providecommand \@ifxundefined [1]{%
 \@ifx{#1\undefined}
}%
\providecommand \@ifnum [1]{%
 \ifnum #1\expandafter \@firstoftwo
 \else \expandafter \@secondoftwo
 \fi
}%
\providecommand \@ifx [1]{%
 \ifx #1\expandafter \@firstoftwo
 \else \expandafter \@secondoftwo
 \fi
}%
\providecommand \natexlab [1]{#1}%
\providecommand \enquote  [1]{``#1''}%
\providecommand \bibnamefont  [1]{#1}%
\providecommand \bibfnamefont [1]{#1}%
\providecommand \citenamefont [1]{#1}%
\providecommand \href@noop [0]{\@secondoftwo}%
\providecommand \href [0]{\begingroup \@sanitize@url \@href}%
\providecommand \@href[1]{\@@startlink{#1}\@@href}%
\providecommand \@@href[1]{\endgroup#1\@@endlink}%
\providecommand \@sanitize@url [0]{\catcode `\\12\catcode `\$12\catcode
  `\&12\catcode `\#12\catcode `\^12\catcode `\_12\catcode `\%12\relax}%
\providecommand \@@startlink[1]{}%
\providecommand \@@endlink[0]{}%
\providecommand \url  [0]{\begingroup\@sanitize@url \@url }%
\providecommand \@url [1]{\endgroup\@href {#1}{\urlprefix }}%
\providecommand \urlprefix  [0]{URL }%
\providecommand \Eprint [0]{\href }%
\providecommand \doibase [0]{http://dx.doi.org/}%
\providecommand \selectlanguage [0]{\@gobble}%
\providecommand \bibinfo  [0]{\@secondoftwo}%
\providecommand \bibfield  [0]{\@secondoftwo}%
\providecommand \translation [1]{[#1]}%
\providecommand \BibitemOpen [0]{}%
\providecommand \bibitemStop [0]{}%
\providecommand \bibitemNoStop [0]{.\EOS\space}%
\providecommand \EOS [0]{\spacefactor3000\relax}%
\providecommand \BibitemShut  [1]{\csname bibitem#1\endcsname}%
\let\auto@bib@innerbib\@empty
\bibitem [{\citenamefont {IBM}()}]{IBMQE}%
  \BibitemOpen
  \bibfield  {author} {\bibinfo {author} {\bibnamefont {IBM}},\ }\href@noop {}
  {}\bibinfo {howpublished}
  {\url{http://www.research.ibm.com/quantum/}}\BibitemShut {NoStop}%
\bibitem [{\citenamefont {Broadbent}\ \emph {et~al.}(2009)\citenamefont
  {Broadbent}, \citenamefont {Fitzsimons},\ and\ \citenamefont
  {Kashefi}}]{broadbent2009universal}%
  \BibitemOpen
  \bibfield  {author} {\bibinfo {author} {\bibfnamefont {A.}~\bibnamefont
  {Broadbent}}, \bibinfo {author} {\bibfnamefont {J.}~\bibnamefont
  {Fitzsimons}}, \ and\ \bibinfo {author} {\bibfnamefont {E.}~\bibnamefont
  {Kashefi}},\ }in\ \href@noop {} {\emph {\bibinfo {booktitle} {Proceedings of
  the 50th Annual IEEE Symposium on Foundations of Computer Science}}}\
  (\bibinfo {organization} {IEEE, New York},\ \bibinfo {year} {2009})\ pp.\
  \bibinfo {pages} {517--526}\BibitemShut {NoStop}%
\bibitem [{\citenamefont {Fitzsimons}\ and\ \citenamefont
  {Kashefi}()}]{fitzsimons2012unconditionally}%
  \BibitemOpen
  \bibfield  {author} {\bibinfo {author} {\bibfnamefont {J.~F.}\ \bibnamefont
  {Fitzsimons}}\ and\ \bibinfo {author} {\bibfnamefont {E.}~\bibnamefont
  {Kashefi}},\ }\href@noop {} {\bibinfo  {journal} {arXiv:1203.5217}\
  }\BibitemShut {NoStop}%
\bibitem [{\citenamefont {Aharonov}\ \emph {et~al.}()\citenamefont {Aharonov},
  \citenamefont {Ben-Or},\ and\ \citenamefont
  {Eban}}]{aharonov2008interactive}%
  \BibitemOpen
\bibfield  {journal} {  }\bibfield  {author} {\bibinfo {author} {\bibfnamefont
  {D.}~\bibnamefont {Aharonov}}, \bibinfo {author} {\bibfnamefont
  {M.}~\bibnamefont {Ben-Or}}, \ and\ \bibinfo {author} {\bibfnamefont
  {E.}~\bibnamefont {Eban}},\ }\href@noop {} {\bibinfo  {journal}
  {arXiv:0810.5375}\ }\BibitemShut {NoStop}%
\bibitem [{\citenamefont {Morimae}\ and\ \citenamefont
  {Fujii}(2012)}]{morimae2012blind}%
  \BibitemOpen
\bibfield  {journal} {  }\bibfield  {author} {\bibinfo {author} {\bibfnamefont
  {T.}~\bibnamefont {Morimae}}\ and\ \bibinfo {author} {\bibfnamefont
  {K.}~\bibnamefont {Fujii}},\ }\href@noop {} {\bibfield  {journal} {\bibinfo
  {journal} {Nat. Commun.}\ }\textbf {\bibinfo {volume} {3}},\ \bibinfo {pages}
  {1036} (\bibinfo {year} {2012})}\BibitemShut {NoStop}%
\bibitem [{\citenamefont {Morimae}\ and\ \citenamefont
  {Fujii}(2013)}]{morimae2013blind}%
  \BibitemOpen
  \bibfield  {author} {\bibinfo {author} {\bibfnamefont {T.}~\bibnamefont
  {Morimae}}\ and\ \bibinfo {author} {\bibfnamefont {K.}~\bibnamefont
  {Fujii}},\ }\href@noop {} {\bibfield  {journal} {\bibinfo  {journal} {Phys.
  Rev. A}\ }\textbf {\bibinfo {volume} {87}},\ \bibinfo {pages} {050301}
  (\bibinfo {year} {2013})}\BibitemShut {NoStop}%
\bibitem [{\citenamefont {Morimae}(2014)}]{morimae2014verification}%
  \BibitemOpen
  \bibfield  {author} {\bibinfo {author} {\bibfnamefont {T.}~\bibnamefont
  {Morimae}},\ }\href@noop {} {\bibfield  {journal} {\bibinfo  {journal} {Phys.
  Rev. A}\ }\textbf {\bibinfo {volume} {89}},\ \bibinfo {pages} {060302}
  (\bibinfo {year} {2014})}\BibitemShut {NoStop}%
\bibitem [{\citenamefont {Hayashi}\ and\ \citenamefont
  {Morimae}(2015)}]{hayashi2015verifiable}%
  \BibitemOpen
  \bibfield  {author} {\bibinfo {author} {\bibfnamefont {M.}~\bibnamefont
  {Hayashi}}\ and\ \bibinfo {author} {\bibfnamefont {T.}~\bibnamefont
  {Morimae}},\ }\href@noop {} {\bibfield  {journal} {\bibinfo  {journal} {Phys.
  Rev. Lett.}\ }\textbf {\bibinfo {volume} {115}},\ \bibinfo {pages} {220502}
  (\bibinfo {year} {2015})}\BibitemShut {NoStop}%
\bibitem [{\citenamefont {Dunjko}\ \emph {et~al.}(2012)\citenamefont {Dunjko},
  \citenamefont {Kashefi},\ and\ \citenamefont {Leverrier}}]{dunjko2012blind}%
  \BibitemOpen
  \bibfield  {author} {\bibinfo {author} {\bibfnamefont {V.}~\bibnamefont
  {Dunjko}}, \bibinfo {author} {\bibfnamefont {E.}~\bibnamefont {Kashefi}}, \
  and\ \bibinfo {author} {\bibfnamefont {A.}~\bibnamefont {Leverrier}},\
  }\href@noop {} {\bibfield  {journal} {\bibinfo  {journal} {Phys. Rev. Lett.}\
  }\textbf {\bibinfo {volume} {108}},\ \bibinfo {pages} {200502} (\bibinfo
  {year} {2012})}\BibitemShut {NoStop}%
\bibitem [{\citenamefont {Hajdu{\v{s}}ek}\ \emph {et~al.}(2015)\citenamefont
  {Hajdu{\v{s}}ek}, \citenamefont {P{\'e}rez-Delgado},\ and\ \citenamefont
  {Fitzsimons}}]{hajduvsek2015device}%
  \BibitemOpen
  \bibfield  {author} {\bibinfo {author} {\bibfnamefont {M.}~\bibnamefont
  {Hajdu{\v{s}}ek}}, \bibinfo {author} {\bibfnamefont {C.~A.}\ \bibnamefont
  {P{\'e}rez-Delgado}}, \ and\ \bibinfo {author} {\bibfnamefont {J.~F.}\
  \bibnamefont {Fitzsimons}},\ }\href@noop {} {\bibfield  {journal} {\bibinfo
  {journal} {arXiv:1502.02563}\ } (\bibinfo {year} {2015})}\BibitemShut
  {NoStop}%
\bibitem [{\citenamefont {Gheorghiu}\ \emph {et~al.}(2015)\citenamefont
  {Gheorghiu}, \citenamefont {Kashefi},\ and\ \citenamefont
  {Wallden}}]{gheorghiu2015robustness}%
  \BibitemOpen
  \bibfield  {author} {\bibinfo {author} {\bibfnamefont {A.}~\bibnamefont
  {Gheorghiu}}, \bibinfo {author} {\bibfnamefont {E.}~\bibnamefont {Kashefi}},
  \ and\ \bibinfo {author} {\bibfnamefont {P.}~\bibnamefont {Wallden}},\
  }\href@noop {} {\bibfield  {journal} {\bibinfo  {journal} {New J. Phys.}\
  }\textbf {\bibinfo {volume} {17}},\ \bibinfo {pages} {083040} (\bibinfo
  {year} {2015})}\BibitemShut {NoStop}%
\bibitem [{\citenamefont {Mantri}\ \emph {et~al.}(2013)\citenamefont {Mantri},
  \citenamefont {P{\'e}rez-Delgado},\ and\ \citenamefont
  {Fitzsimons}}]{mantri2013optimal}%
  \BibitemOpen
  \bibfield  {author} {\bibinfo {author} {\bibfnamefont {A.}~\bibnamefont
  {Mantri}}, \bibinfo {author} {\bibfnamefont {C.~A.}\ \bibnamefont
  {P{\'e}rez-Delgado}}, \ and\ \bibinfo {author} {\bibfnamefont {J.~F.}\
  \bibnamefont {Fitzsimons}},\ }\href@noop {} {\bibfield  {journal} {\bibinfo
  {journal} {Phys. Rev. Lett.}\ }\textbf {\bibinfo {volume} {111}},\ \bibinfo
  {pages} {230502} (\bibinfo {year} {2013})}\BibitemShut {NoStop}%
\bibitem [{\citenamefont {P{\'e}rez-Delgado}\ and\ \citenamefont
  {Fitzsimons}(2015)}]{perez2015iterated}%
  \BibitemOpen
  \bibfield  {author} {\bibinfo {author} {\bibfnamefont {C.~A.}\ \bibnamefont
  {P{\'e}rez-Delgado}}\ and\ \bibinfo {author} {\bibfnamefont {J.~F.}\
  \bibnamefont {Fitzsimons}},\ }\href@noop {} {\bibfield  {journal} {\bibinfo
  {journal} {Phys. Rev. Lett.}\ }\textbf {\bibinfo {volume} {114}},\ \bibinfo
  {pages} {220502} (\bibinfo {year} {2015})}\BibitemShut {NoStop}%
\bibitem [{\citenamefont {Reichardt}\ \emph {et~al.}(2013)\citenamefont
  {Reichardt}, \citenamefont {Unger},\ and\ \citenamefont
  {Vazirani}}]{reichardt2013classical}%
  \BibitemOpen
  \bibfield  {author} {\bibinfo {author} {\bibfnamefont {B.~W.}\ \bibnamefont
  {Reichardt}}, \bibinfo {author} {\bibfnamefont {F.}~\bibnamefont {Unger}}, \
  and\ \bibinfo {author} {\bibfnamefont {U.}~\bibnamefont {Vazirani}},\
  }\href@noop {} {\bibfield  {journal} {\bibinfo  {journal} {Nature (London)}\
  }\textbf {\bibinfo {volume} {496}},\ \bibinfo {pages} {456} (\bibinfo {year}
  {2013})}\BibitemShut {NoStop}%
\bibitem [{\citenamefont {Gheorghiu}\ \emph {et~al.}()\citenamefont
  {Gheorghiu}, \citenamefont {Wallden},\ and\ \citenamefont
  {Kashefi}}]{gheorghiu2015rigidity}%
  \BibitemOpen
  \bibfield  {author} {\bibinfo {author} {\bibfnamefont {A.}~\bibnamefont
  {Gheorghiu}}, \bibinfo {author} {\bibfnamefont {P.}~\bibnamefont {Wallden}},
  \ and\ \bibinfo {author} {\bibfnamefont {E.}~\bibnamefont {Kashefi}},\
  }\href@noop {} {\bibinfo  {journal} {arXiv:1512.07401}\ }\BibitemShut
  {NoStop}%
\bibitem [{\citenamefont {Barz}\ \emph {et~al.}(2012)\citenamefont {Barz},
  \citenamefont {Kashefi}, \citenamefont {Broadbent}, \citenamefont
  {Fitzsimons}, \citenamefont {Zeilinger},\ and\ \citenamefont
  {Walther}}]{barz2012demonstration}%
  \BibitemOpen
\bibfield  {journal} {  }\bibfield  {author} {\bibinfo {author} {\bibfnamefont
  {S.}~\bibnamefont {Barz}}, \bibinfo {author} {\bibfnamefont {E.}~\bibnamefont
  {Kashefi}}, \bibinfo {author} {\bibfnamefont {A.}~\bibnamefont {Broadbent}},
  \bibinfo {author} {\bibfnamefont {J.~F.}\ \bibnamefont {Fitzsimons}},
  \bibinfo {author} {\bibfnamefont {A.}~\bibnamefont {Zeilinger}}, \ and\
  \bibinfo {author} {\bibfnamefont {P.}~\bibnamefont {Walther}},\ }\href@noop
  {} {\bibfield  {journal} {\bibinfo  {journal} {Science}\ }\textbf {\bibinfo
  {volume} {335}},\ \bibinfo {pages} {303} (\bibinfo {year}
  {2012})}\BibitemShut {NoStop}%
\bibitem [{\citenamefont {Barz}\ \emph {et~al.}(2013)\citenamefont {Barz},
  \citenamefont {Fitzsimons}, \citenamefont {Kashefi},\ and\ \citenamefont
  {Walther}}]{barz2013experimental}%
  \BibitemOpen
  \bibfield  {author} {\bibinfo {author} {\bibfnamefont {S.}~\bibnamefont
  {Barz}}, \bibinfo {author} {\bibfnamefont {J.~F.}\ \bibnamefont
  {Fitzsimons}}, \bibinfo {author} {\bibfnamefont {E.}~\bibnamefont {Kashefi}},
  \ and\ \bibinfo {author} {\bibfnamefont {P.}~\bibnamefont {Walther}},\
  }\href@noop {} {\bibfield  {journal} {\bibinfo  {journal} {Nat. Phys.}\
  }\textbf {\bibinfo {volume} {9}},\ \bibinfo {pages} {727} (\bibinfo {year}
  {2013})}\BibitemShut {NoStop}%
\bibitem [{\citenamefont {Fisher}\ \emph {et~al.}(2014)\citenamefont {Fisher},
  \citenamefont {Broadbent}, \citenamefont {Shalm}, \citenamefont {Yan},
  \citenamefont {Lavoie}, \citenamefont {Prevedel}, \citenamefont {Jennewein},\
  and\ \citenamefont {Resch}}]{fisher2014quantum}%
  \BibitemOpen
  \bibfield  {author} {\bibinfo {author} {\bibfnamefont {K.}~\bibnamefont
  {Fisher}}, \bibinfo {author} {\bibfnamefont {A.}~\bibnamefont {Broadbent}},
  \bibinfo {author} {\bibfnamefont {L.}~\bibnamefont {Shalm}}, \bibinfo
  {author} {\bibfnamefont {Z.}~\bibnamefont {Yan}}, \bibinfo {author}
  {\bibfnamefont {J.}~\bibnamefont {Lavoie}}, \bibinfo {author} {\bibfnamefont
  {R.}~\bibnamefont {Prevedel}}, \bibinfo {author} {\bibfnamefont
  {T.}~\bibnamefont {Jennewein}}, \ and\ \bibinfo {author} {\bibfnamefont
  {K.}~\bibnamefont {Resch}},\ }\href@noop {} {\bibfield  {journal} {\bibinfo
  {journal} {Nat. Commun.}\ }\textbf {\bibinfo {volume} {5}},\ \bibinfo {pages}
  {3074} (\bibinfo {year} {2014})}\BibitemShut {NoStop}%
\bibitem [{\citenamefont {Greganti}\ \emph {et~al.}(2016)\citenamefont
  {Greganti}, \citenamefont {Roehsner}, \citenamefont {Barz}, \citenamefont
  {Morimae},\ and\ \citenamefont {Walther}}]{greganti2016demonstration}%
  \BibitemOpen
  \bibfield  {author} {\bibinfo {author} {\bibfnamefont {C.}~\bibnamefont
  {Greganti}}, \bibinfo {author} {\bibfnamefont {M.-C.}\ \bibnamefont
  {Roehsner}}, \bibinfo {author} {\bibfnamefont {S.}~\bibnamefont {Barz}},
  \bibinfo {author} {\bibfnamefont {T.}~\bibnamefont {Morimae}}, \ and\
  \bibinfo {author} {\bibfnamefont {P.}~\bibnamefont {Walther}},\ }\href@noop
  {} {\bibfield  {journal} {\bibinfo  {journal} {New J. Phys.}\ }\textbf
  {\bibinfo {volume} {18}},\ \bibinfo {pages} {013020} (\bibinfo {year}
  {2016})}\BibitemShut {NoStop}%
\bibitem [{\citenamefont {Marshall}\ \emph {et~al.}(2016)\citenamefont
  {Marshall}, \citenamefont {Jacobsen}, \citenamefont {Sch{\"a}fermeier},
  \citenamefont {Gehring}, \citenamefont {Weedbrook},\ and\ \citenamefont
  {Andersen}}]{marshall2016continuous}%
  \BibitemOpen
  \bibfield  {author} {\bibinfo {author} {\bibfnamefont {K.}~\bibnamefont
  {Marshall}}, \bibinfo {author} {\bibfnamefont {C.~S.}\ \bibnamefont
  {Jacobsen}}, \bibinfo {author} {\bibfnamefont {C.}~\bibnamefont
  {Sch{\"a}fermeier}}, \bibinfo {author} {\bibfnamefont {T.}~\bibnamefont
  {Gehring}}, \bibinfo {author} {\bibfnamefont {C.}~\bibnamefont {Weedbrook}},
  \ and\ \bibinfo {author} {\bibfnamefont {U.~L.}\ \bibnamefont {Andersen}},\
  }\href@noop {} {\bibfield  {journal} {\bibinfo  {journal} {Nat. Commun.}\
  }\textbf {\bibinfo {volume} {7}},\ \bibinfo {pages} {13795} (\bibinfo {year}
  {2016})}\BibitemShut {NoStop}%
\bibitem [{\citenamefont {Aaronson}\ \emph {et~al.}()\citenamefont {Aaronson},
  \citenamefont {Cojocaru}, \citenamefont {Gheorghiu},\ and\ \citenamefont
  {Kashefi}}]{aaronson2017implausibility}%
  \BibitemOpen
  \bibfield  {author} {\bibinfo {author} {\bibfnamefont {S.}~\bibnamefont
  {Aaronson}}, \bibinfo {author} {\bibfnamefont {A.}~\bibnamefont {Cojocaru}},
  \bibinfo {author} {\bibfnamefont {A.}~\bibnamefont {Gheorghiu}}, \ and\
  \bibinfo {author} {\bibfnamefont {E.}~\bibnamefont {Kashefi}},\ }\href@noop
  {} {\bibinfo  {journal} {arXiv:1704.08482}\ }\BibitemShut {NoStop}%
\bibitem [{\citenamefont {Gan}\ and\ \citenamefont
  {Harrison}(2005)}]{gan2005calibrating}%
  \BibitemOpen
\bibfield  {journal} {  }\bibfield  {author} {\bibinfo {author} {\bibfnamefont
  {Z.}~\bibnamefont {Gan}}\ and\ \bibinfo {author} {\bibfnamefont {R.~J.}\
  \bibnamefont {Harrison}},\ }in\ \href@noop {} {\emph {\bibinfo {booktitle}
  {Proceedings of the ACM/IEEE SC 2005 Conference}}}\ (\bibinfo {organization}
  {IEEE},\ \bibinfo {year} {2005})\ pp.\ \bibinfo {pages} {22--22}\BibitemShut
  {NoStop}%
\bibitem [{\citenamefont {Senko}\ \emph {et~al.}(2014)\citenamefont {Senko},
  \citenamefont {Smith}, \citenamefont {Richerme}, \citenamefont {Lee},
  \citenamefont {Campbell},\ and\ \citenamefont {Monroe}}]{senko2014coherent}%
  \BibitemOpen
  \bibfield  {author} {\bibinfo {author} {\bibfnamefont {C.}~\bibnamefont
  {Senko}}, \bibinfo {author} {\bibfnamefont {J.}~\bibnamefont {Smith}},
  \bibinfo {author} {\bibfnamefont {P.}~\bibnamefont {Richerme}}, \bibinfo
  {author} {\bibfnamefont {A.}~\bibnamefont {Lee}}, \bibinfo {author}
  {\bibfnamefont {W.}~\bibnamefont {Campbell}}, \ and\ \bibinfo {author}
  {\bibfnamefont {C.}~\bibnamefont {Monroe}},\ }\href@noop {} {\bibfield
  {journal} {\bibinfo  {journal} {Science}\ }\textbf {\bibinfo {volume}
  {345}},\ \bibinfo {pages} {430} (\bibinfo {year} {2014})}\BibitemShut
  {NoStop}%
\bibitem [{\citenamefont {Aaronson}\ and\ \citenamefont
  {Arkhipov}(2014)}]{aaronson2014bosonsampling}%
  \BibitemOpen
  \bibfield  {author} {\bibinfo {author} {\bibfnamefont {S.}~\bibnamefont
  {Aaronson}}\ and\ \bibinfo {author} {\bibfnamefont {A.}~\bibnamefont
  {Arkhipov}},\ }\href@noop {} {\bibfield  {journal} {\bibinfo  {journal}
  {Quantum Inf. Comput.}\ }\textbf {\bibinfo {volume} {14}},\ \bibinfo {pages}
  {1383} (\bibinfo {year} {2014})}\BibitemShut {NoStop}%
\bibitem [{\citenamefont {Shor}(1999)}]{shor1999polynomial}%
  \BibitemOpen
  \bibfield  {author} {\bibinfo {author} {\bibfnamefont {P.~W.}\ \bibnamefont
  {Shor}},\ }\href@noop {} {\bibfield  {journal} {\bibinfo  {journal} {SIAM J.
  Comput.}\ }\textbf {\bibinfo {volume} {41}},\ \bibinfo {pages} {303}
  (\bibinfo {year} {1999})}\BibitemShut {NoStop}%
\bibitem [{\citenamefont {Lu}\ \emph {et~al.}(2007)\citenamefont {Lu},
  \citenamefont {Browne}, \citenamefont {Yang},\ and\ \citenamefont
  {Pan}}]{lu2007demonstration}%
  \BibitemOpen
  \bibfield  {author} {\bibinfo {author} {\bibfnamefont {C.-Y.}\ \bibnamefont
  {Lu}}, \bibinfo {author} {\bibfnamefont {D.~E.}\ \bibnamefont {Browne}},
  \bibinfo {author} {\bibfnamefont {T.}~\bibnamefont {Yang}}, \ and\ \bibinfo
  {author} {\bibfnamefont {J.-W.}\ \bibnamefont {Pan}},\ }\href@noop {}
  {\bibfield  {journal} {\bibinfo  {journal} {Phys. Rev. Lett.}\ }\textbf
  {\bibinfo {volume} {99}},\ \bibinfo {pages} {250504} (\bibinfo {year}
  {2007})}\BibitemShut {NoStop}%
\bibitem [{\citenamefont {Lanyon}\ \emph {et~al.}(2007)\citenamefont {Lanyon},
  \citenamefont {Weinhold}, \citenamefont {Langford}, \citenamefont {Barbieri},
  \citenamefont {James}, \citenamefont {Gilchrist},\ and\ \citenamefont
  {White}}]{lanyon2007experimental}%
  \BibitemOpen
  \bibfield  {author} {\bibinfo {author} {\bibfnamefont {B.}~\bibnamefont
  {Lanyon}}, \bibinfo {author} {\bibfnamefont {T.}~\bibnamefont {Weinhold}},
  \bibinfo {author} {\bibfnamefont {N.}~\bibnamefont {Langford}}, \bibinfo
  {author} {\bibfnamefont {M.}~\bibnamefont {Barbieri}}, \bibinfo {author}
  {\bibfnamefont {D.}~\bibnamefont {James}}, \bibinfo {author} {\bibfnamefont
  {A.}~\bibnamefont {Gilchrist}}, \ and\ \bibinfo {author} {\bibfnamefont
  {A.}~\bibnamefont {White}},\ }\href@noop {} {\bibfield  {journal} {\bibinfo
  {journal} {Phys. Rev. Lett.}\ }\textbf {\bibinfo {volume} {99}},\ \bibinfo
  {pages} {250505} (\bibinfo {year} {2007})}\BibitemShut {NoStop}%
\bibitem [{\citenamefont {Kiesel}\ \emph {et~al.}(2005)\citenamefont {Kiesel},
  \citenamefont {Schmid}, \citenamefont {Weber}, \citenamefont {Ursin},\ and\
  \citenamefont {Weinfurter}}]{kiesel2005linear}%
  \BibitemOpen
  \bibfield  {author} {\bibinfo {author} {\bibfnamefont {N.}~\bibnamefont
  {Kiesel}}, \bibinfo {author} {\bibfnamefont {C.}~\bibnamefont {Schmid}},
  \bibinfo {author} {\bibfnamefont {U.}~\bibnamefont {Weber}}, \bibinfo
  {author} {\bibfnamefont {R.}~\bibnamefont {Ursin}}, \ and\ \bibinfo {author}
  {\bibfnamefont {H.}~\bibnamefont {Weinfurter}},\ }\href@noop {} {\bibfield
  {journal} {\bibinfo  {journal} {Phys. Rev. Lett.}\ }\textbf {\bibinfo
  {volume} {95}},\ \bibinfo {pages} {210505} (\bibinfo {year}
  {2005})}\BibitemShut {NoStop}%
\bibitem [{\citenamefont {Reichardt}\ \emph {et~al.}()\citenamefont
  {Reichardt}, \citenamefont {Unger},\ and\ \citenamefont
  {Vazirani}}]{reichardt2012classical}%
  \BibitemOpen
  \bibfield  {author} {\bibinfo {author} {\bibfnamefont {B.~W.}\ \bibnamefont
  {Reichardt}}, \bibinfo {author} {\bibfnamefont {F.}~\bibnamefont {Unger}}, \
  and\ \bibinfo {author} {\bibfnamefont {U.}~\bibnamefont {Vazirani}},\
  }\href@noop {} {\bibinfo  {journal} {arXiv:1209.0448}\ }\BibitemShut
  {NoStop}%
\bibitem [{Sup()}]{Supplemental}%
  \BibitemOpen
\bibfield  {journal} {  }\href@noop {} {\bibinfo  {journal} {See Supplemental
  Material for more information about the security analysis, and the details
  for CNOT gates, CHSH games, state tomography, process tomography}\
  }\BibitemShut {NoStop}%
\bibitem [{\citenamefont {Wang}\ \emph {et~al.}(2016)\citenamefont {Wang},
  \citenamefont {Chen}, \citenamefont {Li}, \citenamefont {Huang},
  \citenamefont {Liu}, \citenamefont {Chen}, \citenamefont {Luo}, \citenamefont
  {Su}, \citenamefont {Wu}, \citenamefont {Li}, \citenamefont {Lu},
  \citenamefont {Hu}, \citenamefont {Jiang}, \citenamefont {Peng},
  \citenamefont {Li}, \citenamefont {Liu}, \citenamefont {Chen}, \citenamefont
  {Lu},\ and\ \citenamefont {Pan}}]{PhysRevLett.117.210502}%
  \BibitemOpen
\bibfield  {journal} {  }\bibfield  {author} {\bibinfo {author} {\bibfnamefont
  {X.-L.}\ \bibnamefont {Wang}}, \bibinfo {author} {\bibfnamefont {L.-K.}\
  \bibnamefont {Chen}}, \bibinfo {author} {\bibfnamefont {W.}~\bibnamefont
  {Li}}, \bibinfo {author} {\bibfnamefont {H.-L.}\ \bibnamefont {Huang}},
  \bibinfo {author} {\bibfnamefont {C.}~\bibnamefont {Liu}}, \bibinfo {author}
  {\bibfnamefont {C.}~\bibnamefont {Chen}}, \bibinfo {author} {\bibfnamefont
  {Y.-H.}\ \bibnamefont {Luo}}, \bibinfo {author} {\bibfnamefont {Z.-E.}\
  \bibnamefont {Su}}, \bibinfo {author} {\bibfnamefont {D.}~\bibnamefont {Wu}},
  \bibinfo {author} {\bibfnamefont {Z.-D.}\ \bibnamefont {Li}}, \bibinfo
  {author} {\bibfnamefont {H.}~\bibnamefont {Lu}}, \bibinfo {author}
  {\bibfnamefont {Y.}~\bibnamefont {Hu}}, \bibinfo {author} {\bibfnamefont
  {X.}~\bibnamefont {Jiang}}, \bibinfo {author} {\bibfnamefont {C.-Z.}\
  \bibnamefont {Peng}}, \bibinfo {author} {\bibfnamefont {L.}~\bibnamefont
  {Li}}, \bibinfo {author} {\bibfnamefont {N.-L.}\ \bibnamefont {Liu}},
  \bibinfo {author} {\bibfnamefont {Y.-A.}\ \bibnamefont {Chen}}, \bibinfo
  {author} {\bibfnamefont {C.-Y.}\ \bibnamefont {Lu}}, \ and\ \bibinfo {author}
  {\bibfnamefont {J.-W.}\ \bibnamefont {Pan}},\ }\href@noop {} {\bibfield
  {journal} {\bibinfo  {journal} {Phys. Rev. Lett.}\ }\textbf {\bibinfo
  {volume} {117}},\ \bibinfo {pages} {210502} (\bibinfo {year}
  {2016})}\BibitemShut {NoStop}%
\bibitem [{\citenamefont {Hofmann}(2005)}]{hofmann2005complementary}%
  \BibitemOpen
  \bibfield  {author} {\bibinfo {author} {\bibfnamefont {H.~F.}\ \bibnamefont
  {Hofmann}},\ }\href@noop {} {\bibfield  {journal} {\bibinfo  {journal} {Phys.
  Rev. Lett.}\ }\textbf {\bibinfo {volume} {94}},\ \bibinfo {pages} {160504}
  (\bibinfo {year} {2005})}\BibitemShut {NoStop}%
\end{thebibliography}%

\end{document}